\documentclass[a4paper,twocolumn,11pt]{quantumarticle}

\pdfoutput=1
\usepackage[utf8]{inputenc}
\usepackage[english]{babel}
\usepackage[T1]{fontenc}

\usepackage{amsmath,amssymb, amsthm}

\usepackage{tikz}
\usepackage{lipsum}
\usepackage{physics}
\usepackage{bbm}

\usepackage{mathtools}
\usepackage{hyperref}
\usepackage{cleveref}
\usepackage{lipsum}

\newtheorem{theorem}{Theorem}

\usepackage[numbers,sort&compress]{natbib}

\begin{document}

\title{Efficient learning of sparse Pauli-Lindblad models for fully connected qubit topology}

\author{Jose Este Jaloveckas}
\affiliation{Infineon Technologies AG, Am Campeon 1-15, Neubiberg, Germany}
\affiliation{Technische Universität München, Munich, Germany}
\affiliation{Ludwig Maximilian Universität, Munich, Germany}
\author{Minh Tham Pham Nguyen}
\affiliation{Infineon Technologies AG, Am Campeon 1-15, Neubiberg, Germany}
\author{Lilly Palackal}
\affiliation{Infineon Technologies AG, Am Campeon 1-15, Neubiberg, Germany}
\affiliation{Technische Universität München, Munich, Germany}
\author{Jeanette Miriam Lorenz}
\affiliation{Ludwig Maximilian Universität, Munich, Germany}
\affiliation{Fraunhofer Institute for Cognitive Systems IKS, Munich, Germany}

\author{Hans Ehm}
\affiliation{Infineon Technologies AG, Am Campeon 1-15, Neubiberg, Germany}

\maketitle

\begin{abstract}
The challenge to achieve practical quantum computing considering current hardware size and gate fidelity is the sensitivity to errors and noise. Recent work has shown that by learning the underlying noise model capturing qubit cross-talk, error mitigation can push the boundary of practical quantum computing. This has been accomplished using Sparse Pauli-Lindblad models only on devices with a linear topology connectivity (i.e. superconducting qubit devices). In this work we extend the theoretical requirement for learning such noise models on hardware with full connectivity (i.e. ion trap devices). 
\end{abstract}

\section{Introduction}
\label{introduction}
The central promise of quantum computing is to enable algorithms that can provide both polynomial and super-polynomial speed-ups over the best-known classical algorithms for a special set of problems. Practical use of these algorithms would require millions of two-qubit gates. However, current state-of-the-art implementations of two-qubit gates across various platforms (superconducting, trapped-ion, quantum dots) still suffer from error rates around $10^{-2}$ to  $10^{-3}$ \cite{Huang2020, Bruzewicz2019, Burkard2023}, rendering any solutions generated from these quantum algorithms unreliable or even completely unusable. Thus, for quantum computers to successfully solve real-world problems, it is necessary to address the challenge of noise: reducing the errors which occur in elementary physical components due to unwanted or imperfect interactions. 

There are currently three paradigms to tackle noise: quantum error suppression \cite{Viola1998, lvarez2010, Pokharel2018, Biercuk2009_1, Biercuk2009_2}, quantum error mitigation ~\cite{Li2017, Temme2017, vandenBerg2023, Kim2023, Filippov2023} and quantum error correction ~\cite{Shor1995, Preskill1998, Kitaev1997, Aharonov1997, Gottesman1998}. Error suppression intends to improve the coherence time of noisy hardware via optimal control and operates close to the hardware level. On the other end of the spectrum, quantum error correction intends to achieve fault tolerant quantum computing by detecting and inverting individual faults during the computation, effectively creating noiseless hardware. Even though the theory of quantum error correction can provide an answer for fault tolerant quantum computing in the long term, at the moment, it is not feasible experimentally on a large scale \cite{Fowler2012, Campbell2017} due to the large quantum control overhead, error rates above the allowed critical thresholds and low quantum volumes of current Noisy Intermediate-Scale Quantum (NISQ) era hardware. To fill the gap between error suppression and error correction, error mitigation does not improve the coherence time, but rather utilises post processing techniques operated by classical computers to recover an unbiased estimation of relevant expectation values. 

Recently, \textit{Quek et. al} \cite{Eisert2022} have shown that error mitigation may perform worse than previously expected at circuit depth of $O(poly(log(n))$. Despite this, in practical implementations, IBM researchers have recently showcased positive outcomes through the utilisation of Probabilistic Error Cancellation (PEC) and Zero-Noise Extrapolation (ZNE) with sparse Pauli–Lindblad (SPL) models \cite{vandenBerg2023, Kim2023}, on superconducting quantum processors with fixed-frequency transmon qubits and topology patterns of heavy-hexagon lattices, by employing circuits that do not exceed the coherence time of the qubits. In \cite{Filippov2023} error mitigation techniques based on tensor networks and SPL models are used to reduce the overhead of PEC. 

Previous work \cite{Eisert2020} and these latest results have brought to light that effective error mitigation techniques must heavily rely on information of the underlying noise. On this matter SPL models allow for the efficient characterization of cross-talk between qubits, thus opening the door to error mitigation to be applied in practical scenarios. In particular \cite{vandenBerg2023, Kim2023} use cycle-benchmarking to characterize two qubit Clifford gate noise in a superconducting qubit hardware device with linear topology circuits, as depicted in Fig. 2 and Fig. 4 of \cite{vandenBerg2023}. So far the qubit topologies considered for learning SPL models are solely of this type for two main reasons: (1) superconducting qubit hardware devices are the go-to platform due to their availability, and (2) it allows to maintain a constant number of prepared circuits during noise characterization, as elucidated in Theorem SIV.4 in the supplementary material of \cite{vandenBerg2023}.  

For the particular application considered in \cite{vandenBerg2023, Kim2023}, (a trotterized evolution of an Ising chain model) linear qubit topology suffices to avoid introducing SWAP gates. This is not the case in general and in many algorithms of interest a linear topology would introduce a large SWAP gate overhead which directly increases computation time, introducing decoherence and thus deteriorating the results. To avoid this, one option is to use quantum hardware with fully connected qubit topologies. Ion trap quantum computers have shown promise in scalability and gate fidelity and have been shown to perform better for certain applications \cite{https://doi.org/10.48550/arxiv.2305.07092}. For this reason it is of interest to scale the methods presented in \cite{vandenBerg2023, Kim2023} to fully connected qubit topologies. Moreover the use of cycle benchmarking has been proposed as the base for a layer fidelity metric \cite{McKay2023} to benchmark the performance of quantum processors and avoid the use of quantum volume \cite{Cross2019} which scales poorly with increasing number of qubits. 

In \Cref{sec:preliminaries} we introduce SPL models (following \cite{vandenBerg2023}) and cycle benchmarking for noise characterization (following \cite{vandenBerg2023} and \cite{Chen2023}). In \Cref{sec:main_results} we re-state the theorem presented in \cite{vandenBerg2023} and we generalize it to fully connected qubit topologies. 

\section{Preliminaries}\label{sec:preliminaries}

We start by considering the SPL model proposed by \textit{Van den Berg et. al} in \cite{vandenBerg2023}. We model an $N$-qubit Pauli noise channel $\Lambda$ that arises from a sparse set of local interactions, according to a Lindblad Master equation \cite{Breuer2007} with generator $\mathcal{L}(\rho) = \sum_{k\in\mathcal{K}}\lambda_k(P_k\rho P_k^\dagger-\rho)$, where $\mathcal{K}$ represents a set of local Paulis $P_k=\sigma_{k_1}\otimes \cdots \otimes \sigma_{k_N}$ supported on $N$ qubits and $\lambda_k$ denotes the corresponding model coefficients. The resulting noise model is then given by (see \cite{vandenBerg2023})

\begin{equation}
    \Lambda(\rho)=\text{exp}[\mathcal{L}](\rho) = \prod_{k\in\mathcal{K}}\Big(w_k \rho + (1-w_k) P_k\rho P_k^\dagger\Big) \,,
\end{equation}
where $w_k = 2^{-1}(1+e^{-2\lambda_k})$. The local Paulis in $\mathcal{K}$ are chosen to reflect the noise interactions in the quantum processor. Their number, which determines the model complexity and expressivity, typically scales polynomially in $N$ and therefore allows us to represent noise models for the full device by a small set of non-negative coefficients $\lambda_k$. We define the weight pattern of a Pauli string $P_k$ supported on $n$ qubits to be an $N$-bit binary string with 0 indicating identity and 1 indicating non-identity. The weight of the Pauli string is the element-wise sum of the weight pattern, and we denote the weight of the Pauli $P_k$ by $w_{pt}(P_k)$. 

We learn the set of coefficients $\lambda_k$ by applying cycle benchmarking (CB) \cite{Erhard2019,Chen2023}. This technique estimates the fidelity of a set of relevant Paulis $P_k$ by repetition of the same noise process $d$ times and by measuring the corresponding expectation value at each depth. The fidelity can then be extracted from the decay rates in the resulting curves and the model coefficients fitted from this corresponding Pauli fidelity. Details on CB  are given in \cite{Breuer2007, Chen2023}. 

 A naive first approach to apply CB prepares a set of depth-increasing circuits for each Pauli of interest in $\mathcal{K}$ and measures the circuit in the same basis as the prepared Pauli. This becomes intractable even for polynomially increasing sets of model Pauli strings. However, given a single basis it is possible to estimate a large number of fidelities using the same measurements. By considering a k-local SPL noise model, only low-weight Pauli strings are included in $\mathcal{K}$, i.e. $P_k\in \mathcal{K}$, such that $w_{pt}(P_k)=k< N$. Then it suffices to consider all of the $\{X,Y,Z\}^{\otimes k}$ bases on each set of $k$ qubits. A natural choice is to include in $\mathcal{K}$ only two-local Paulis, \textit{i.e.} Pauli string with weight equal to 1 or 2, between pairs of qubits where a two-qubit gate can be directly implemented, which results in a small set $\mathcal{K}$ and captures cross-talk between qubits. From this point forward let us consider this type of two-local SPL models.

We describe the qubit topology of the quantum hardware as a undirected graph $G=(V,E)$ with a set of vertices $v_i\in V$ corresponding to physical qubits and edges $(v_i,v_j)\in E$ if a two-qubit gate can be directly implemented between qubits $v_i$ and $v_j$. We consider each measurement basis as an $n$ elements Pauli string. The qubit topology bounds the number of Pauli string such that for each $(v_i, v_j )\in E$ the substrings at locations $v_i$ and $v_j$ exactly cover $\{X, Y, Z\}^{\otimes 2}$. Considering linear qubit topologies, \textit{Van den Berg et. al} showed in \cite{vandenBerg2023} that it is suffiant to use a total of nine Pauli strings independently of the number of qubits. In the next section we re-state their theorem proving this statement and generalize it to the case of fully connected qubit topologies. For this we use the notion of the maximum clique of a graph.  A clique of graph $G$ is a subset of vertices $C\subseteq V$ such that every pair of vertices in $C$ are adjacent. The maximum clique of the graph is the clique with highest cardinality. If there is no case for confusion, we sometimes refer to maximum clique of a graph as the cardinality of the maximum clique.

\section{Main results}\label{sec:main_results}%

We now generalise the result of Theorem SIV.4 from \cite{vandenBerg2023} to fully connected qubit topologies. Let us first restate it verbatim for clarity.

\begin{theorem}{Theorem SIV.4 from \cite{vandenBerg2023}.}
    Given a qubit topology represented by a graph $(V,E)$ whose vertices are ordered in such a way that no vertex $v \in V$ is preceded by more than two connected vertices. Then there exist nine Pauli strings  such that for each $(v_i, v_j )\in E$ the substrings at locations $v_i$ and $v_j$ exactly cover $\{X, Y, Z\}^{\otimes 2}$.
    \label{theorem:SIV vandenBerg2023}
\end{theorem}

The proof of \Cref{theorem:SIV vandenBerg2023} as presented in \cite{vandenBerg2023} relies on traversing the graph and reducing the proof to the 2 and 3 connected components. Here we extend \Cref{theorem:SIV vandenBerg2023} by considering the maximum clique of the graph and then generalize it to fully connected graphs. We first show that the same result as  \Cref{theorem:SIV vandenBerg2023} of a constant number can be achieved by limiting the maximum clique of the graph to 4. This is a less restrictive constraint than those outlined in \Cref{theorem:SIV vandenBerg2023}.

\begin{theorem}
    Given a qubit topology represented by a graph $(V,E)$ with maximum clique size  at most 4. Then there exist nine Pauli strings such that for each $(v_i, v_j )\in E$, the substrings at locations $v_i$ and $v_j$ exactly cover $\{X, Y, Z\}^{\otimes 2}$. 
    \label{theorem:reformulation SIV}
\end{theorem}

\begin{figure}[ht]
    \centering
    \includegraphics[width = 6cm]{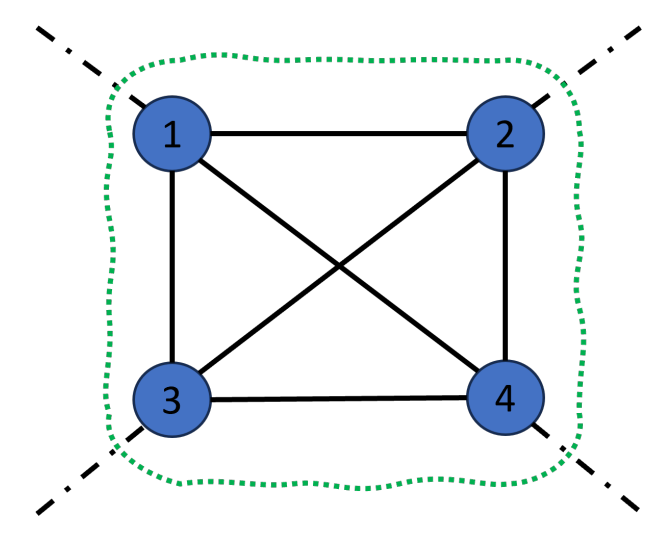}
    \caption{A clique of size 4 which is part of a larger graph $G$. We label the vertices in the clique from 1 to 4.
}
    \label{fig:maxclique4}
\end{figure}

\begin{proof}
 Given a graph $G = (V,E)$ with clique size at most 4. Let us consider a clique of size 4 as part of the graph $G$ as depicted in \Cref{fig:maxclique4}. We label the vertices to be 1, 2, 3, 4. We can then consider the nine different 4-Pauli strings given in \Cref{table:9measbasis}. We denote them by $B_k$ for $k=1,\dots,9$. These nine Pauli strings exactly cover $\{X, Y, Z\}^{\otimes 2}$ at positions of any two vertices $v_i$ and $v_j$, where $(v_i, v_j)$ is an edge inside a clique of size 4.

\begin{table}[ht]
\centering
\caption{Each row is a Pauli string representing a measurement bases. }
\begin{tabular}{|c|c|c|c|c|}
\hline
 & Vertex 1 & Vertex 2  & Vertex 3  & Vertex 4  \\ \hline
$B_1$ &X & X & X & X \\ \hline
$B_2$ &X & Y & Y & Y \\ \hline
$B_3$ &X & Z & Z & Z \\ \hline
$B_4$ &Y & X & Z & Y \\ \hline
$B_5$ &Y & Y & X & Z \\ \hline
$B_6$ &Y & Z & Y & X \\ \hline
$B_7$ &Z & X & Y & Z \\ \hline
$B_8$ &Z & Y & Z & X \\ \hline
$B_9$ &Z & Z & X & Y \\ \hline
\end{tabular}
\label{table:9measbasis}
\end{table}

Note that for any clique smaller than 4 we just remove one vertex such that covering $\{X, Y, Z\}^{\otimes 2}$ is still satisfied. If we now consider the entire graph $G$, any cliques of size 4 will follow the pattern described by \Cref{table:9measbasis} and we can extend the Pauli strings $B_k$ to more vertices using the same row-wise pattern. This concludes the proof \Cref{theorem:reformulation SIV}.
\end{proof}

The results in \Cref{theorem:SIV vandenBerg2023} and \Cref{theorem:reformulation SIV} are limited to graphs with clique size at most 4. Our goal is to remove this condition and consider fully connected graphs, as such qubit topologies are already experimentally realized in trapped-ion quantum computers. In \Cref{theorem:full_connectivity} we give an upper bound for the number of Pauli substrings needed in fully connected topology. 

\begin{theorem}
    Given a fully connected qubit topology represented by the graph $K_N$, where $N\geq4$ is the number of qubits. There exists an arrangement which requires $3(1 + 2\lceil log_2( N - 2)\rceil)$ Pauli strings such that for each $(v_i, v_j )\in E$, the Pauli substrings at locations $v_i$ and $v_j$ cover $\{X, Y, Z\}^{\otimes 2}$.
    \label{theorem:full_connectivity}
\end{theorem}

\begin{figure}[ht]
    \centering
    \includegraphics[width=7.5cm]{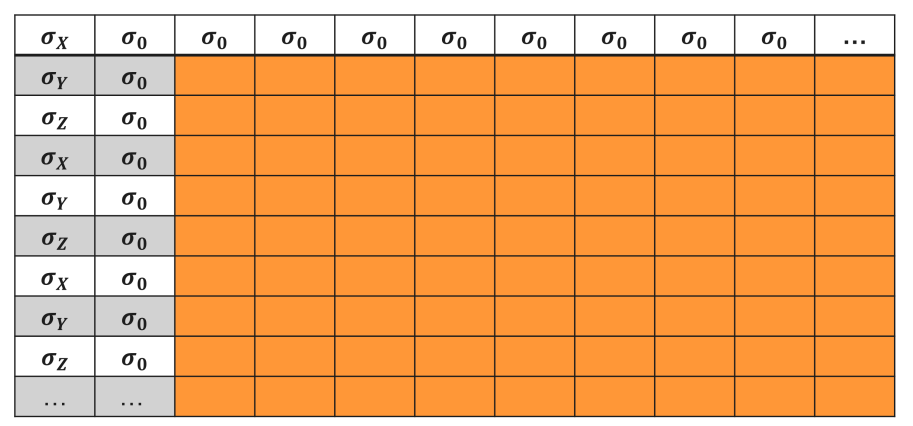}
    \caption{Table representation of assignment of Pauli strings to each qubit. There is one column for each qubit and one row for each three-tuple fo Pauli strings. Without loss of generality, we can assume that the first column will be filled by repeating the pattern [$\sigma_X$, $\sigma_Y$, $\sigma_Z$]. We fill the second column and first row by $\sigma_0$. The only area in which we have not filled (orange area) is from row 2 to beyond, and from column 3 to column $N$.}
    \label{fig:table_proof_theorem3}
\end{figure}

\begin{proof}
    Let us first introduce some notation. As before we will construct a table where rows correspond to the Pauli strings representing a measurement basis and columns correspond to Pauli strings for each vertex. We define the following column substrings:  
    
    \begin{equation}
    \begin{split}
        &\sigma_X = \begin{bmatrix} X \\ X\\ X\end{bmatrix}, \sigma_Y = \begin{bmatrix} Y \\ Y\\ Y\end{bmatrix}, \sigma_Z = \begin{bmatrix} Z \\ Z\\ Z\end{bmatrix},\\
        &\sigma_0 = \begin{bmatrix} X \\ Y\\ Z\end{bmatrix}, \sigma_1 = \begin{bmatrix} Y \\ Z\\ X\end{bmatrix}, \sigma_2 = \begin{bmatrix} Z \\ X\\ Y\end{bmatrix},\\
    \end{split}
    \end{equation}

    Given a fully connected qubit topology $K_N$, where $N\geq 4$ is the number of qubits.

We construct a table as depicted in \Cref{fig:table_proof_theorem3}, where each column $j$ lists Pauli substrings acting on node $j$ and each row defines three measurement bases. For the first column, we only use $\sigma_X$, $\sigma_Y$, $\sigma_Z$. On the second column, we only use $\sigma_0$. Thus, between column 1 and column 2, Pauli substrings of the rows cover $\{X, Y, Z\}^{\otimes 2}$. This can be easily verified checking the rows of the first two columns of \Cref{table:9measbasis}. Now beside cell $(1,1)$ assigned to $\sigma_X$, we fill up the remaining cells of the first row with $\sigma_0$. Thus, the rows from column 2 to column $N$ contain substrings which cover $\{XX, YY, ZZ\}$. We are left to fill the orange area of \Cref{fig:table_proof_theorem3} (column 3 to column $N$), for which we only use $\sigma_1$ and $\sigma_2$. We can use a divide-and-conqueror approach, as show in \Cref{fig:devide_conquer}, to make sure that any pair of columns $(i,j)$ with $i,j \in \{3, \dots, N\}$ contains substrings $\{\sigma_1\odot \sigma_2, \sigma_2 \odot \sigma_1\}$, where $\odot$ denotes the element-wise tensor product of column substrings. This ensures that column 1 and column $k \geq 3$ exactly cover $\{X, Y, Z\}^{\otimes 2}$. Now, it remains to show that substrings between column 2 and any column $k\geq3$  cover $\{X, Y, Z\}^{\otimes 2}\setminus\{XX, YY, ZZ\}$. We can observe that in the following combination, the rows will cover the required space.

\begin{equation}
\begin{split}
      &\{\sigma_0 \odot \sigma_1, \sigma_0 \odot \sigma_2\} \text{ covers } \\
      &\{X, Y, Z\}^{\otimes 2}\setminus \{XX, YY, ZZ\}.
\end{split}
\end{equation}

\begin{figure}[ht]
    \centering
    \includegraphics[width = 8cm]{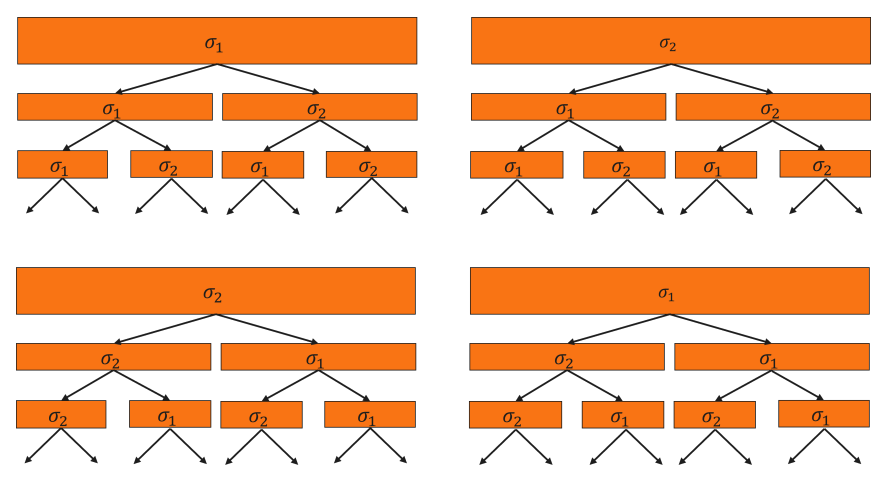}
\caption{Devide-and-conquer approach for assigning Pauli substring in \Cref{fig:table_proof_theorem3}. For the first row of the orange area, we partition the columns to 2 groups. We fill the first group with $\sigma_1$ and the second group by $\sigma_2$. For the next row, we partition each group from the first row into two again, filling them with $\sigma_1$ and $\sigma_2$. We repeat process until we cannot divide it anymore. Up until this point, between any pair of columns, there is a substring $\sigma_1\odot \sigma_2$. We then repeat the same procedure but swapping $\sigma_1 \leftrightarrow \sigma_2$. Finally, we obtain substrings $\sigma_1 \odot \sigma_2$ between any pair of columns.
}
    \label{fig:devide_conquer}
\end{figure}

This concludes the assignment of substrings in \Cref{fig:table_proof_theorem3} such that for each $(v_i, v_j )\in E$ for $N\geq 4$, the Pauli substrings at locations $v_i$ and $v_j$ cover $\{X, Y, Z\}^{\otimes 2}$. We can now turn to see the cost in rows and see that this corresponds to the necessary number of Pauli strings. By counting the tree branches using the divide-and-conquer approach, it is easy to verify that this arrangement requires $1 + 2\lceil log_2( N - 2)\rceil$ rows, i.e. $3\left(1 + 2\lceil log_2( N - 2)\rceil\right)$ Pauli strings.     
\end{proof}

\section{Conclusions and outlook}\label{sec:discussion}
Our work demonstrates that cycle-benchmarking for two-local SPL models can be an efficient approach to error mitigation as the number of required measurements scales logarithmically with the number of qubits. Additionally, we explicitly construct the Pauli strings corresponding to the measurement bases, using a total of $3(1 + 2\lceil log_2( N - 2)\rceil)$ strings. Even though it is trivial to prove that a constant number of 9 measurement bases is insufficient when increasing the number of qubits, it remains an open problem whether our construction corresponds to the minimum number of measurement bases needed.

This work represents a promising first step towards efficient learning of SPL models in ion trap quantum computers or other devices with all-to-all qubit connectivity. More broadly, this technique could be applied to any characterization or measurement scheme of a 2-local observable.

In future work we look forward to explore the practical implementation of this scheme and asses the viability of SPL models for ion-trap hardware architectures. 



\section*{Acknowledgements}
This project is supported by the Federal Ministry for Economic Affairs and Climate Action on the basis of a decision by the German Bundestag through the project Quantum-enabling Services and Tools for Industrial Applications (QuaST). 

\bibliographystyle{quantum}
\bibliography{main}





\end{document}